# Allocating User Requests to Licenses in the Open Mobile Alliance DRM System

Nikolaos Triantafyllou[1], Petros Stefaneas[2] and Panayiotis Frangos[1]

[1] School of Electrical and Computer Engineering, National Technical University of Athens, Heroon Polytechniou 9, 15780 Zografou, Athens, Greece
{nitriant, pfrangos}@central.ntua.gr

[2] School of Applied Mathematical and Physical Sciences, National Technical University of Athens, Heroon Polytechniou 9, 15780 Zografou, Athens, Greece
petros@math.ntua.gr

**Abstract.** We claim that the Open Mobile Alliance (OMA) Order of Rights Object Evaluation algorithm causes the loss of execution permissions under certain circumstances. By introducing an algebraic characterization, as well as an ordering on licenses, we redesign this algorithm to minimize possible errors, in a way suitable for the low computational powers of mobile devices. Our algorithm does not suggest major changes of the existing system, hence its future implementation will have minimal development cost. In addition we provide a formal proof that our algorithm minimizes the above problem. The proof is conducted using the OTS/CafeOBJ method for verifying liveness properties.

## 1 Introduction

Open Mobile Alliance (OMA) is an organization responsible for the definition of standards for the Mobile DRM systems [1]. The proposed standards for the DRM systems are called OMA-DRM [2], while the language, in which licenses are written, is called OMA-REL [3]. Like most such languages [4, 5, 6] OMA-REL is XML based and is defined as a mobile profile of ODRL [2]. OMA's DRM is currently implemented in most mobile devices and smart phones and is adopted by most vendors for mobile content. We demonstrate that the OMA License Allocation Algorithm currently in use suffers from a loss of execution permissions (or rights) and we present a new algorithm to overcome this. We point out here that although this algorithm is designed to address problems of the OMA DRM system, the principals described in this paper (labeling licenses etc.) can be applied to other DRM domains as well, to counter the same problem and allow for an automated license selection with minimal loss of rights.

Barth and Mitchel [7] characterize the loss of rights by introducing a notion of monotonicity in licenses. A license in their notation is said to be monotonic, if a sequence of actions allowed under one license is also allowed under a more flexible set of licenses. By more flexible, we mean a set of licenses that contains the same rights for the contents as the original license, possibly more rights for them with different constraints, and in addition contains rights about different contents. They show that all algorithms that assign actions to rights, as actions occur, are non–monotonic in the context of OMA. They also prove that the problem of determining whether a sequence of actions complies with a license is an NP-complete problem. Monotonicity can be achieved by allowing the DRM agents to revise their past allocation of actions to rights after learning which other actions the consumers wish to perform.

Here we present an algorithm that overcomes most cases of loss of rights without the need to design complex and computationally heavy algorithms for DRM agents that reallocate actions to rights, but instead only a labeling of the licenses is required. There exist some cases where our algorithm is not fully monotonic, but this is intentional. In the case were some form of loss is inevitable, we prompt the user on his preferences to what rights will be lost. This can cause the algorithm to behave non-monotonically, but we believe that in this special case an algorithm that respects the desires of the user is preferable to a fully monotonic algorithm.

Our paper is organized as follows: section 2 presents the algorithm and the loss of execution rights, section 3 gives a brief introduction to order sorted algebra and presents our new algorithm with some case studies. Section 4 introduces the reader to the concepts of Observation Transition Systems (OTS) and the algebraic specification language CafeOBJ. In addition in section 4 a specification of a DRM agent using this new algorithm is provided and the section concludes with a formal proof that our algorithm does not suffer from this loss of rights. Finally section 5 concludes the paper.

## 2   The OMA Choice Algorithm

In a DRM system users may end up with licenses from different sources that serve different purposes, for example advertisement licenses etc. It is possible for these licenses to refer to the same content. So a problem rises as to what license should be considered optimal. OMA answers this by specifying a set of rules that the DRM agent must apply when automatically selecting which rights to apply when accessing content, in the case there are multiple rights for it this content [3]. These rules are in essence an ordering on the constraints defined by OMA REL which are transferred to the constraint sets of permissions. Although not explicitly stated these rules are inevitably projected to the licenses themselves, so for a fixed user request, applying these rules produces an ordering on the licenses themselves. We will refer to these rules as the OMA Allocation Algorithm. For example consider the following two licenses;

*License 1; you can listen to song A or D for one month after the first time you execute the play permission*

*License 2; you can listen to songs A or C for one month starting October 2011*

In the above set of licenses we are posed with a question; what license should the DRM agent choose when requested to play song A? Based on the rules imposed by OMA REL, the DRM agent must select the second license because it contains a date-time constraint *(for one month starting October 2011)* which is ranked higher in the ordering than the interval constraint (*one month after the first execution*) of the first license. The intent behind this is that the interval will start whenever the user chooses while the date-time constraint will expire even if it is not exercised. So it is clear that this rule was created to benefit the user. Indeed when multiple constraints containing a date exist, the one that ends nearest to the present is to be preferred; the other rights can still be used later. In fact all of the rules set out in the specification are of similar intent; unconstraint rights are to be preferred over constraint and so on [3].

So we can argue that the intent of this algorithm, is to allow for an automated decision making process that will result in protecting the interests of the user by choosing to use, the license that maximizes the rights available to him after the execution of a right, although this is not explicitly stated in [3].

## 2.1 The OMA Allocation Algorithm

The OMA allocation algorithm deals with the problem of multiple licenses that refer to the same content. A license is to be preferred if it contains a constraint of higher ranking in this order. It is described as follows [3]:
- Only the rights that are valid at the given time should be taken into account from the algorithm.
- Rights that are *not constrained* should always be preferred over constrained rights.
- Any right that includes a *date-time* constraint, and possibly others, should be preferred over rights that are constrained but do not have such a restriction.
- If there exists more than one rights with a *date-time* constraint, the one with the further in the future *end* element should be preferred.
- If there exist a choice between many rights and none of them contains a *date-time* constraint the one containing an i*nterval* constraint should be preferred if there exists such.
- Rights that contain a c*ount* constraint should be preferred after rights that contain *a timed-count* constraint.

## 2.2 Losing Rights

As we argued, this algorithm is designed to protect the interests of the user, by selecting licenses that allow for a maximum execution of rights under the installed set of licenses. Despite its intent there exist some special cases where using it may in fact lead to a more limited set of execution rights. This is caused mainly because this algorithm does not take into account the future requests of the user. A set of installed licenses $ls$ allows a finite power set of rights, $rights(ls) = \{r_1, ..., r_n\}$, meaning that it is allowed for $r_i, r_j \in rights(ls)$, $i \neq j$ and $r_i = r_j$. We denote by $remnants(ls, l, r)$ the rights that still hold after the execution of a permission matching a request r from a license $l \in ls$. We define as *a loss of rights* the case where $rights(ls) \backslash remnants(ls, l, r) \supset \{r\}$. This denotes that more rights than the request are no longer available.

For user request r, we desire an algorithm that selects a license l containing $r_i$ (matching the request r), such that for all other licenses l' that also contain a right matching r to hold that remnants(ls, l′, r) ⊆ remnants(ls, l, r). We will call this property *minimal loss property*.

To understand how such a loss of rights can occur, assume a set of licenses and the original algorithm. For a user request r about content A, assume the algorithm selects license $l_i$. But this choice may affect the set of rights available for content B, depleting all rights referring to it. In the decision process of the original algorithm we do not take into account content B at all. We are only interested on what is the optimal right in regard to content A. This lack of generality is what causes this *loss of rights*. For a concrete example, let us consider the following two licenses installed in the device of a user.

*License1: "you may listen to songs A or B once before the end of the month"*
*License2: "you may listen to songs A or D ten times."*

These licenses translated in to OMA- REL contain the following constraints: License 1 contains a date-time constraint *("before the end of the month")*, while License 2 contains a count constraint *("ten times")*. If the user decides to use the right "listen to song A", the OMA-DRM agent will select License 1 since it contains a date-time constraint that is ranked higher in the ordering. But by doing so, License 1 will become depleted since it also contains the count constraint denoted by "once". This results in the user losing the right of ever listening to song B with this set of licenses. So, $nants(ls, L1, r) \subseteq remnants(ls, L2, r)$. This would not occur if the agent had selected License 2. This clearly can be seen as against the best interest of the user, which as we argued is the intent of this algorithm and in [7] is characterized as an infuriating situation. Now consider a second example:

*License 1: "you can listen to songs A or B once".*
*License 2: "you can listen to songs A or C or D once".*

Assume that there exists a user request to listen to song A. In order to satisfy the request a loss of rights is inevitable, since all the licenses that contain the request cause a loss. How should an algorithm decide in this case? One option would be to simply allocate the request to the license that causes the minimal loss of rights. But it is possible that the user values the right to listen to song B higher than the right to listen to songs C or D combined. So we claim that in the case that a loss is inevitable the final decision should rest on the user. Instead of the minimal loss property we believe that an allocation algorithm should enjoy a weakened version of it: "*when a loss of rights is not inevitable the minimal loss property holds*". We will refer to this property as the *weak minimal loss property*.

## 3 Redesigning the Algorithm

### 3.1 Order Sorted Algebra in a Nutshell

An order sorted algebra is a partial ordering ≤ on a set of sorts [11], where by sorts we usually mean a set of names for data types. This subsort relation imposes a restriction on an S-sorted algebra A, by s-sorted algebra we mean a mapping between the sort names and sub sets from the set A called the carriers of sort s, that if $s \leq s'$ then $A_s \subseteq A_{s'}$, where $A_s$ denotes the elements of sort s in A. Order sorted algebra (OSA) [11] provides a way for several forms of polymorphism and overloading, error definition, detection and recovery, multiple inheritance, selectors when there are multiple constructors, retracts, partial operations made total on equationally defined sub-sorts, an operational semantics that executes equations as left-to-right rewrite rules and many more applications [11].

Formally, given a partially ordered sort set S, an *S-sorted set* A is just a family of sets $A_s$ for each sort $s \in S$. A many-sorted *signature* is a pair $(S, \Sigma)$ where S is called the sort set and $\Sigma$ is an $S^* \times S$-sorted family of functions $\{\Sigma_{w,s} | w \in S^* \text{ and } s \in S\}$. An *order sorted signature* is a triple $(S, \leq, \Sigma)$ such that $(S, \Sigma)$ is a many-sorted signature, $(S, \leq)$ is a poset, and the operations satisfy the following monotonicity condition; $\sigma \in \Sigma_{w1,s1} \cap \Sigma_{w2,s2}$ and $w1 \leq w2$ imply $s1 \leq s2$. Given a many-sorted signature an $(S, \Sigma)$-algebra A is a family of sets $\{A_s | s \in S\}$ called the carriers of A, together with a function $A_\sigma: A_w \to A_s$ for each σ in $\Sigma_{w,s}$, where $A_w = A_{s1} \times ... \times A_{sn}$, $w=s_1..s_n$.

### 3.2 Licenses in OMA REL/DRM

A license in OMA REL is a set of constructs we will call sublicenses. A sublicense is defined as a set of constraints, that when met, authorize a set of constraint permissions. A constraint permission now, is defined as a set of constraints that when met authorize in turn a set of permissions, or rights. Permission are defined as an action on a content that is protected by the DRM system. The actions defined by OMA REL are; *play, display, print, execute and export.* After allowing an action the DRM agent transforms the license by updating the constraints. So licenses in OMA REL resemble in their structure a forest of trees where several sublicenses constitute a license. The structure of a sublicense can be seen in figure 1. The idea behind this structure is that a set of permissions is allowed if its "local" set of constraints is satisfied and the sublicenses constraints are met as well.

Based on the ideas of order sorted algebra, we can argue that licenses are basically a data type. Consecutively there exist a sort set S, of sort names that can be used to represent these licenses. In addition, based on the rights object evaluation provided by the OMA Allocation Algorithm there exists a predefined ordering on these sorts. So if we identify the order sorted algebra that is "hidden" in the definition of licenses we can create an algorithm. The basic idea is to simply apply this ordering to decide what the most suitable license to use is.

The key observation that leads to the redesign of the algorithm is that there exist some special cases that cause the loss of rights. Given a set of licenses, we can lose rights when: *"A license contains more than one permission elements and after the execution some part of it becomes depleted"*. According to this statement, we can argue that each license should be characterized by the following observations:

    A. Some part of the license becomes depleted after the execution of a right
    B. The license contains more than one permission elements
    C. The characterizing constraint based on the OMA constraint ordering

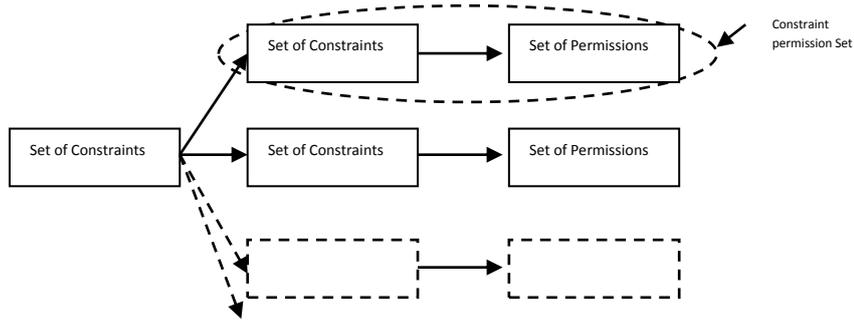

Figure 1. A sublicense in OMA REL

In order for A to occur, the license needs to contain either a count constraint (with one execution still allowed), or a timed count constraint (with once again one execution left). B can be easily checked at the level of the creation of a license. Finally, C can be achieved by a simple search on the license constraints.

We can now define an order sorted signature for OMA REL licenses as ($S_1 \times S_2 \times S_3$, $\leq$, $\Sigma$) with $\Sigma = \emptyset$, $S_3$ = {Count, Timed-count, Date-time, Interval, True} the names of the various constraints allowed by OMA REL where we omit the constraints; individual, system, accumulated because they do not play any part in the decision made by the original algorithm. $S_1$ = {Once, Many} denoting whether the license will allow more than one execution of its permissions. This can be easily checked, by first observing if the license contains a count or timed-count constraint. These constraints contain a positive integer stating the number of times the right can be executed and the DRM agent must reduce this number with each execution (in the case of timed-count the reduction occurs only after the passing of a specified time limit as well). Finally $S_2$ = {Complex, Simple}, denoting if the license contains more than one permissions.

So essentially a license in this approach is an element of the carrier set $A_s$ for each sort s where $s \in s_1 \times s_2 \times s_3$. For example $Once \times Simple \times Count$, denotes that the license allows only one more execution of a right, it contains only one right and the dominant constraint of the license is a count constraint. The ordering comes from the predefined ordering of the rights object evaluation algorithm in conjunction with the definintions: $once \leq many$ and $simple \leq complex$. So, formally we have that $s_1 \times s_2 \times s_3 \leq s'_1 \times s'_2 \times s'_3$ implies that $s_1 < s'_1$ or ($s_1 = s'_1$ and $s_2 < s'_2$ or ($s_2 = s'_2$ and $s_3 \leq s'_3$)). Using this ordering on licenses we will present in the next section a new algorithm for the decision problem of the optimal license.

Our approach does not require any knowledge on behalf of the agent on the future actions of the users. It uses the existing OMA allocation algorithm in its core and its computational weight is moved to the creation of the licenses. Once this is done, the algorithm we propose is no more than a linear search on the licenses and an application of the initial OMA algorithm. We believe that our algorithm is relatively easy to implement since it does not require the reconstruction of all the DRM agents.

### 3.3 Labeling Licenses

We could augment a license to contain those sort names by using labels that will be added to the sublicenses as a top label and to the constraint permission sets as a local label. This should be done simultaneously with the creation of the licenses to reduce the computational cost on the mobile devices. We want each of these labels to denote when it comes to a constraint permission set or a sublicense:
a) If it contains one or more permission elements/constraint permission sets; we define that the label will contain the designated word "*Simple*" or "*Complex*" respectively.
b) The optimal constraint Name (based on the OMA ordering), we define that the label will contain the name of the Constraint, e.g. "*Count*", "*Datetime*", "*True*" (for unconstraint) and so on.
c) Whether it becomes depleted after the next execution of one of its rights. Here we define that the label will contain the words "*Once*" or "*Many*" respectively.

As an example, such a label for a constraint permission set could be "*Simple × Many × True*". In addition we assume that the DRM agent is enhanced so as to be able to update these labels after the execution of permissions as necessary. Meaning that if a license after the execution of permission allows only one more execution its label should be updated to say "*Simple×Once×True*" from "*Simple×Many×True*".

Formally a license enhanced with these labels is defined as a set of sublicenses. Each sublicense, sub-l, is a triplet $< Cons, CPerm, label >$ where Cons is a set of constraints, CPerm is a set of constraint permission sets, and finally label is a label. We define operators to retrieve them as *Constraints (sub-l), CPS (sub-l)* and *label (sub-l)* respectively. Now each constraint permission set, CP, contains three elements as well, a set of constraints, a set of permissions and a label, $< Cons, Perm, label >$. We define operators to retrieve these as: *Constraints(CP), Perm(CP)* and *label(CP)* respectively. For a CP now we define that, $label\ (CP) = times\ \times\ Simple\ \times\ constraint$ if $\#(Perm(CP)) = 1$. This means that a set of constraint permissions is labeled as $times\ \times\ Simple\ \times\ constraint$ if and only if it contains only one permission. Else it is labeled $es\ \times\ Complex\ \times\ constraint$. Where $times$ and $constraint$ are variables with $times \in S_2$ and $constraint \in S_3$. Now $label\ (CP) = Once\ \times\ complexity\ \times\ constraint$ if the execution of any one of the permissions in Perm(CP) causes the Constraints(CP) to no longer hold. This means that CP can only be used one more time. Else it is labeled as $Many\ \times\ complexity\ \times\ constraint$. Where complexity is a variable, $complexity \in S_1$. A sublicense now is labeled as, $label\ (sub - l)\ = times \times Simple \times constraint$ if $\#\ (CPS\ (sub - l)) = 1$. This means that the sub-license contains only one constraint permission set. Else it is labeled $times\ \times Complex\ \times constraint$. Finally a sublicense is labeled as, $label\ (sub - l) = Once\ \times\ complexity\ \times constraint$ if the execution of a permission belonging to any of the constraint permission sets of the sub-license, CPS(sub-l), causes Constraints(sub-l) to no longer hold. Else $label(sub - l) = Many\ \times\ complexity\ \times\ constraint$. For example consider the sublicense: sub-l = {Once before the end of the month} either {up to ten times play songs A or B} or {once print document C}. Here we have that Constraint(sub-l) ={ one time, before the end of the month}. CPS(sub-l) = { CP1, CP2 } where CP1 = { (up to ten times) ( play songs A, play song B)} and CP2 = {(one time) (print document C)}. So here #(CPS(sub-l)) = 2.

### 3.3 The proposed new Algorithm

Now that we have defined the labels for the sublicenses/constraint permission sets we can present the proposed algorithm. Informally the intuition behind it is: *"check the licenses installed on the mobile DRM device for the ones matching the request of the user and see which do not fall into the special cases. If all fall into that category prompt the user. Else, if there exists a set of licenses that do not fall on this special category use the OMA Allocation Algorithm on them to obtain the optimal license. Finally update the labels if necessary".*

For label $L = complexity \times times \times constraint$, we define $times(L) = times$, $comp(L) = complexity$ and $cons(L) = constraint$. For a constraint permission set CP and a request r, we define $sat(CP,r) = true$ if there exists a permission in CP that satisfies r. Similarly for a sublicense subl, we define $sat(subl,r) = true$ if there exists a $CP \in subl$ such that $sat(CP,r)$. For a license l $sat(l,r) = true$ if there exists $subl \in l$ such that $sat(subl,r)$. Finally for a set of licenses ls, $sat(ls,r) = true$ if there exists $l \in ls$ such that $sat(l,r)$. An abstract version of the algorithm is shown in table 1.

In the following sections we formally verify that this algorithm satisfies the weakened minimal loss property. This property guarantees that the algorithm will provide the user with the maximal possible set of available rights after the satisfaction of each request and at the same time respect his preferences of rights when a loss is inevitable. The cornerstone idea behind this algorithm is that a loss of rights will occur when the selected license contains the user request in a constraint permission set, CP, of a sublicense, subl, such that either $label(subl) = Complex \times Once \times Constraint$ or if $label(subl) \neq Complex \times Once \times Constraint$ then $label(CP) = Complex \times Once \times Constraint$. The goal of the algorithm is to avoid such selections if that is possible. Else if only such licenses match the request, the user is prompted to ensure the satisfaction of his preference on the available rights.

The first stage of the selection produces a set of licenses whose sublicenses that contain the request will either not get depleted after the actions execution or they contain only one set of constraint permissions. On the second stage the selection leaves us with a set of licenses that, on a constraint permission level, will either not get depleted after the execution or contain only one permission element, the permission matching the user request.

Table 1. An abstract representation of the algorithm

```
1. Input: a user request r, a set of licenses ls
2. Store all licenses l ∈ ls for which sat(r,ls) holds. If
   only one such license exists return that license.
3. Else, store all licenses, l, from step 2, for which it
   holds that sat(subl,r) ∧ subl ∈ l ∧ not (times(label(subl)) =
   once ∧ comp(label(subl)) = complex).
4. Search the licenses from step 3, and store all licenses
   l, such that sat(CP,r) ∧ CP ∈ subl ∧ subl ∈ l for which not
   (times(label(CP)) = once ∧ compl(label(CP)) = complex ).
5. If the result of step 4 is not the empty set, then run
```

```
         the original OMA allocation algorithm on only those
         licenses and return the result.
  6.     Else prompt the user to select a license from step 2.
```

Denoting as *<label(subl),label( CP)>* the pairs of labels for the sublicense and constraint permission set respectively, the algorithm only selects pairs of the following form; (i)*<complexity × Many × constraint, Simple × times × constraint>, (ii) <complexity ×Many × constraint, complexity × Many X constraint>, (iii) <Simple × times × constraint, Simple × times × constraint>* and *(iv) <Simple × times × constraint, complexty × Many × constraint>*. So if the set from the second stage exists, it will contain the licenses matching the request that become depleted if and only if they only contain a single permission matching request and nothing else. Using the above algorithm there exist only two ways for a permission to get lost. The first case is when all the licenses for which $sat(l,r)$ cause a loss. In this case the user is prompted by the algorithm as to which rights he prefers to loose. So in this case it is clear that the algorithm protects the preferences of the users and we consider this loss as intentional. The second case occurs when there exists only one license l such that $sat(l,r)$ and l causes a loss of rights. But as we must always try to satisfy the request if there exists a suitable license, this loss is considered intentional as well.

### 3.4 Case Studies

We have implemented the above algorithm in Java and were able to do several case studies so as to check that no loss of rights occurs. In Table 2 we can see some of the license sets checked. As we can see in the Table 2 the two algorithms agree on licenses where the special cases do not occur while the proposed algorithm correctly chooses the license that will not cause any loss of permissions in the case of these special circumstances.

Table 2. Case Studies implemented in Java

| License Set | User request | Proposed Algorithm | OMA Choice Algorithm |
|---|---|---|---|
| License 1: *"play songs A or B once before the end of the month"* <br> License 2*: "play songs A or C ten times"* | Play song A | License 2 | License 1 |
| License 1: "for *a total of ten times, display content 1 or until the end of the month play content2"* <br> License 2: *"until the end of the month play content 2 three times or display content 1".* | Display content 1 | License 2 | License 2 |
| License 1: "for *a total of ten times, display content 1 or until the end of the month play content2"* <br> License 2: *"until the end of the month play content 2 three times or display content 1".* | Play content 2 | License 2 | License 2 |
| License 1: "for *a total of ten times, display content 1 or play content2"* <br> License 2: *"play content 2 three times or display content 1".* <br> License 3: *"play content 2 or 3 once before the end of the month"* | Play content 2 | License 1 | License 3 |

We believe that the algorithm proposed here better captures the intent of the OMA specification. The originals algorithms intent is to allow for automated selection, while allowing for the maximum possible permissions being available to the user. Tthe original algorithm fails in doing so for some special cases, on the other hand our algorithm is designed to overcome these losses of rights, without altering the original algorithm significantly.

## 4. Verification of the proposed algorithm

Case studying can never replace a formal proof no matter how thorough. In this section we present a formal proof that the proposed algorithm satisfies the weakened minimal loss property. The proof was conducted by specifying an arbitrary OMA DRM system that uses this algorithm, as an Observation Transition System (OTS) [12]. Next we defined this OTS in CafeOBJ notation [13], an algebraic specification langue that allows computer human interactive theorem proving.

We wish to prove that the *weakened minimal loss property* holds. We assume a set of licenses $ls$, installed in a DRM agent such that for each request r, if there exist a set of licenses ls' for which $\forall l \in ls'$, $sat(l,r)$ the algorithm always selects $l \in ls'$, else nothing is selected. There exist three cases.

a) Only one license exists such that $sat(l,r)$. For the property to hold the algorithm in this case must select l.
b) All the licenses for which $sat(l,r)$ cause a loss of rights. Here the property holds trivially. This is the case where the algorithm prompts the user.
c) There exists $l \in ls$ such that $sat(l,r)$ and l does not cause a loss of rights. In order for an algorithm to satisfy the weakened minimal loss property here one of the following must hold:
  1. If there exists l such that $sat(l,r)$ and no part of l gets depleted then l must be selected
  2. If $c_1$) does not hold then, there exists a license l', such that $hts(ls) \setminus remants(ls, l', r) = \{r\}$. For the property to hold $l'$ must be selected.

Indeed in case $c_1$) we have that $remants(ls, l, r) = rights(ls)$, so for all other $l' \in ls$ such that $sat(l', r)$ we will have $remants(ls, l', r) \subseteq rights(ls)$, so the property will hold. Likewise in $c_2$) we have that $rights(ls) \backslash remants(ls, l', r) \supseteq \{r\}$. So if there exists l such that $sat(l, r)$ and $rights(ls) \backslash remants(ls, l, r) = \{r\}$ then for all other l' that $sat(l', r)$ it will hold that $remants(ls, l', r) \subseteq remants(ls, l, r)$.

It is trivial to prove that our algorithm selects l such that $sat(l, r)$ since it is explicitly stated in step two that the selection takes place from only these licenses. Point $c_1$) is also trivial since when there exists a license that will not get depleted then both the sublicense, sub-l, and the constraint permission set, CP, for which $sat(sub-l, r)$ and $sat(CP, r)$ holds we have that $times(sub-l) = Many$ and $times(CP) = Many$. The algorithm explicitly states in step three and four that it selects such licenses.

What remains to be verified for the proposed algorithm is the disjunction of bullets (a), (b) and ($c_2$). We will call this disjunction Property 3.

### 4.1 Connection between weakened minimal loss property and liveness property

In order to verify Property 3 we will transform it to the liveness property of figure 2 and verify that instead. We assume a set of licenses $ls$ and that for a requests r the algorithm selects $l \in ls$ such that $sat(l, r)$, as discussed there is no need to verify this. Next we define a coloring on the permissions of $rights(ls)$. Initially all the permissions are colored white. If no part of l is depleted after the execution of the permission corresponding to r, then the coloring of the permissions remains unchanged. If some part of $l$ gets depleted, causing the rights $rights(ls) \backslash remnants(ls, l, r)$ to become unavailable, we define that the color of $p \in \{rights(ls) \backslash remnants(ls, l, r)\}$ becomes black, if l is the only license such that $sat(l, r)$ or if all the other licenses $l'$ for which $sat(l', r)$ holds also cause a loss of rights or if p matches the user request r. In all other cases the color of p remains unchanged.

**Proposition**. Assume that for all permissions $p \in rights(ls)$, the user makes a request r that matches p an infinite number of times (fairness), and that no license gets depleted by any other means except the selection and use of some permission. Finally, since when there exists l such that $sat(l, r)$ and no part of l gets depleted, it is straightforward to prove that l is selected by our algorithm; we can assume that there are no such licenses in ls. Formally the latter is expressed as; for all licenses $l \in ls$, all sublicenses $sl \in l$, and all constraint permission sets $cp \in CPS(sl)$, if $time(label(sl)) = Many$, then $time(label(cp)) = Once$. Under these three assumptions we claim that $Property 4 \rightarrow Property 3$.

| *Property 4* |
|---|
| *if a permission right belongs to the installed licenses of the DRM system and is colored White* that *leads-to it being colored Black* |

Figure 2. Liveness property that corresponds to the weakened minimal loss property

**Proof.** Due to the third assumption when selected all licenses from ls have some part of them depleted. So after each request $rights(ls)\backslash remnants(ls,l,r) \neq \emptyset$. For the liveness property to hold it suffices to prove that after the satisfaction of a request all the unavailable permissions will get colored black. Indeed the coloring of $p \in rights(l)$ is defined to only change when $p \in rights(ls)\backslash remnants(ls,l,r)$. Since Property 4 claims that all the permissions get colored black if they belong to $rights(ls)$ and originally all permissions are colored white, the change can only occur when $p \in rights(ls)\backslash remnants(ls,l,r)$. So for the liveness property to hold, after each selection all the rights of $rights(ls)\backslash remnants(ls,l,r)$ must get colored black. But that will only happen if there exists only one license such that $sat(l,r)$ or when all other licenses $l' \neq l$ for which $sat(l',r)$ cause a loss of rights as well or when $rights(ls)\backslash remnants(ls,l,r) = \{r\}$, i.e. when Property 3 holds. So the liveness property holds only when Property 3 holds, i.e., Property 4 implies Property 3. This is the property we will verify for the OTS in CafeOBJ specification of the algorithm.

### 4.2 Observation Transition Systems, Behavioral Objects and CafeOBJ

An Observation Transition System (OTS) is a transition system that can be written in terms of equations. Assuming a universal state space, say *Y*, an OTS *S* is a triplet *S* = <*O, I, T*> where *I* is a subset of *Y*, the set of initial states of the system and *O* is a set of observation operators. Each observer in O is a function that takes a state of the system and possibly a series of other data type values (visible sorts) and returns a value of a data type that is characteristic to that state of the system. Given an OTS *S* and two states $u_1, u_2 \in Y$, the equivalence $(u_1 =_S u_2)$ between them wrt *S* is defined as $\forall o \in O, o(u_1) = o(u_2)$. The previous equality creates the equivalence classes, $Y/=_S$, on the states of an OTS. Finally, *T* is the set of transition, conditional functions (or actions). Each transition takes as input a state of the system and again possibly a series of data-type values and returns a new state of the system. If $\tau \in T$ then $\tau(u_1) = \tau(u_2)$ for each $u_1, u_2 \in Y/=_S$. For each $u \in Y, \tau(u)$ is called the successor state of u wrt τ. The condition $c_\tau$ of $\tau$ is called the effective condition. Also, for each $u \in Y, \tau(u) = u$ if $\neg c_\tau(u)$.

An OTS defines a *Behavioral Object*, *Behavioral Object Composition* methodology has been defined formally in [14, 15]. From the state of the composite object we can retrieve the state of the component objects via *Projection Operators* [14, 15]. There are several ways to compose an object from component objects. *Parallel Composition* without Synchronization, if the changes on the states of an object do not affect the states of the other objects of the same level. *Parallel Composition* with *Synchronization* when the changes in the state of one object may alter the state of an object in the same level. In respect to the number of objects that compose a composite object, we have *Dynamic Composition* if that number of component objects is not fixed. Else the composition is called Static.

CafeOBJ is an algebraic specification language [13]. The choice was made because an OTS can be written in CafeOBJ in a natural way. Moreover, hierarchical behavioral object composition, has already been defined in [14, 15] with the use of

CafeOBJ. The universal state space $Y$ is denoted in CafeOBJ by a hidden sort, while each observer by an observation operator. Assuming visible sorts $V_{ij}$ and $V$ that correspond to the data types $D_k$ and $D$, where $k = i_1, \ldots, i_m$, the observation operator denoting $o_{i1,\ldots,im}$ is declared as follows; *bop o* : $V_{i1} \ldots V_{im} H \rightarrow V$. Any state in $I$ is denoted by a constant, say *init*, which is declared as: *op init* : $\rightarrow H$. A transition $\tau_{j1,\ldots,jn} \in T$ is denoted by a CafeOBJ action operator: *bop* $\tau$ : $V_{j1} \ldots V_{jm} H \rightarrow H$, with $V_k$ a visible sort corresponding to the data type $D_k$ and $k = j_1, \ldots, j_n$. Each transition is defined by defining the value returned by each observer in the successor state, when $\tau_{j1,\ldots,jn}$ is applied in a state $u$. When $c - \tau_{j1,\ldots,jn}(u)$ holds, this is expressed generally by a conditional equation denoted by the keyword *ceq*. The value returned by $o_{i1,\ldots,im}$ is not changed if $\tau_{j1,\ldots,jn}$ is applied in a state $u$ such that $\neg(c - \tau_{j1,\ldots,jn}(u))$. The basic building blocks of a CafeOBJ specification are modules. Each module defines a sort. CafeOBJ provides built in modules for the most commonly used data-types (visible sorts) like BOOL (that specifies propositional logic), and so on. An underscore _ in the definition of an operator indicates the place where an argument is put. For example the definition of the negation operator could be `op not_ : Bool -> Bool`. The pervious states that not is an operator (denoted by the keyword `op`) that after the string `not` takes as input a visible value of the sort Bool and returns a value of the visible sort Bool again. The keyword `mod!` (`mod*`) indicates that the module being defined has tight (loose) semantics. Visible (hidden) sorts are denoted enclosing them within `[and]` (`*[and]*`). The keyword `eq` is used to denote an equation and as already mentioned `ceq` to denote a conditional equation. Modules can be imported using the keyword `pr`. Finally the key word `bop` is used to denote observation and action operators (with `bops` being used to denote more than one simultaneously).

### 4.3 Proving Liveness properties in CafeOBJ

The OTS/CafeOBJ method has been successfully applied to the verification of systems safety properties, by many researchers and the industry including our group [8-10]. The property we must prove here is a liveness property, which is different than safety properties. It is a temporal property of the form: *something good will eventually happen*. The procedure to prove such properties is briefly explained below and the reader is invited to consult [16] for a full review of the subject. In addition we refer the readers interested in the verification of safety properties to [12-13]. Finally we should point out here that this is one of the first attempts to apply this technique to verify a real algorithm and this holds in, our opinion, some merit of its own, as it also displays the power of this technique.

#### 4.3.1 Liveness Properties

Predicates whose type is *Y:* → *Bool* are called *state predicates*. Every state predicate in this paper does not have any quantifiers unless explicitly stated. Also we must point out that free variables in formulas in this setting are assumed to be universally quantified. As described in more detail in [16] there exist five kind of basic properties in the OTS/CafeOBJ framework, inspired by UNITY [17]. We assume that $p, q, r, p_j,$ are arbitrary state predicates, $J$ an arbitrary set and $u, u'$ two reachable states of the system. Given an OTS $S$ the five basic properties are defined, $p\ unless\ q$, $stable_S p$, $invariant_S q$, $p\ ensures\ q$ and $p\ leads-to_S q$ (or $p \mapsto_S q$).

Informally $p\ unless_s q$ states that whenever predicate p holds in a state and q does not hold then after applying a transition rule we get to a state that either p will hold or q. $Stable_s p$ states that once we reach a state that p holds p keeps holding, although reaching such a state is not guaranteed. $Invariant_s q$ says that q holds in all reachable states of the OTS S. Next $p\ ensures_s q$ states that if S reaches a state where $p$ holds, S will eventually reach a state where $q$ holds. Finally $p\ leads\text{-}to_s q$ is similar to ensure but here $p$ is not guaranteed to hold until $q$ becomes true. Also leads-to properties are transitive from definition while ensures are not. In [16] a more thorough analysis of the above informal is presented.

### 4.3.2 Verification with CafeOBJ

The literature on the verification of safety (invariant) properties using the OTS/CafeOBJ *Proof Score* method is quite extensive [13,8,9,10,18,19] and many more. Readers are invited to refer to the above literature as the verification of liveness properties usually requires the verification of lemmas, which most likely will be invariant (safety) properties. Here we will describe briefly how to verify that *S* satisfies *ensures* and *leads-to* properties. This together with the verification of *invariant* properties covers all of the five basic properties we described above because proving *ensures* properties consists of proving *unless* properties and finally *stable* properties are special cases of unless properties. Suppose that the OTS S is specified in a module say SYS, which defines a hidden sort H. Also we assume that the invariant properties wrt S are declared in a module say INV.

Proving a property of the form $p\ ensures_s q$, is equivalent with the proof of $p\ unless_s q\ \land p\ eventually_s q$. The first predicate is called the "unless" case while the second the "eventually" case.

In order to prove the "unless" case we must show that for all transition rules the unless inductive step is preserved. For an arbitrary state s and an arbitrary successor state s' this step informally states that if p holds in s and q does not hold in s implies that in s' either p or q will hold. In CafeOBJ notation this will look something like:

```
eq ustep(z_a) = (unl_p(s, z_a) and not (unl_q(s, z_a))) implies
(unl_p(s', z_a) or unl_q(s', z_a)) .
```

Where `unl_p` and `unl_q` are CafeOBJ formulas representing predicates p and q respectively. If these are defined in a module say UNL and for a transition rule t we prove that t preserves the unless step with a proof score like the following:

```
open USTEP
-- Declaration of constants
op y_1 : -> V_1 .
...
op y_n : -> V_n .
```

```
Declaration of equations denoting case_i
-- declaration of the successor state
eq s'= t(s, y₁, ..., yₙ) .
-- check if the predicate holds
red Lemmas implies ustep(z₁, …, zₐ) .
close
```

Above $y_i$ are constants denoting arbitrary values of the data types. Also `red` is a CafeOBJ command used for the reduction of the given term based on the equations defined in the open module. Finally Lemmas is a CafeOBJ term constructed by the conjunction of some predicates used as lemmas, to discard cases that return false. Notice here that since the variables $z_i$ occur freely in p and q they are equivalent to being universally quantified. If CafeOBJ returns *true* in all sub-cases and all transition rules, the proof of the "*unless case*" is successfully completed.

For the "*eventually case*" the inductive step informally states that if in s p holds and q does not hold then q will hold in a successor state s'. Here we must prove that there exists a witness of a transition that makes the eventual step true. If that rule is t then it can be checked with a proof like the previous by replacing the unless step with a term say `estep(z₁, …, zₐ)` that defines the eventually step in CafeOBJ notation and opening the module where this is defined.

Rewriting is used to verify that an OTS satisfies leads-to properties, based on a set of deduction rules defined in [16]. A technicality is that we cannot use the logical operators that are defined in the built-in CafeOBJ module Bool, this is analyzed in [16]. So we need to re-define them and define an operator to evaluate them. This is done in a new module, usually called OTSLOGIC. For example the evaluation of and can be defined as: `eq eval (P/\Q) = eval(P) and eval(Q) .` Where P and Q are CafeOBJ variables. Next, since reasoning for leads-to properties requires invariant, ensures and lead-to properties, we must also define the appropriate operators in OTSLOGIC, `invariant, ensures` and `|-->` respectively.

Finally we must define the equations that correspond to the conditional deductive rules as they are defined in [16]. In [16] three kind of such parameterized rules are defined. The parameters can be instantiated with the operators `invariant, ensures` and `|-->`. This gives a total of 21 conditional rules that defined in `OTSLOGIC` that we can use to reason about leads-to properties.

**4.4 Specification of the proposed algorithm as an OTS in CafeOBJ**

**4.4.1 Basic Data Types**

Before introducing the OTSs that specify a DRM agent that implements the proposed algorithm, we need first specify the data types that are required. These specifications consist of *modules* that define *visible sorts*, each of whom corresponds to a data type. In order to specify the OTSs the following data types where required:

1. *Content*, defines the visible sort `cont`, and specifies the abstract concept of a digital content.
2. *Permission*, defines the visible sorts `action`, `perm`. It imports the module

```
        mod* Label{ [type1 , type2 ,type3 < label]
        op _=_ : label label -> Bool {comm}
        op _=_ : type1 type1 -> Bool {comm}
        op _=_ : type2 type2 -> Bool {comm}
        op _=_ : type3 type3 -> Bool {comm}
        ops simple complex  : -> type1
        ops count datetime true : -> type2
        ops once many : -> type3
        op _ & _ & _ : type1 type2 type3 -> label
        op type1?_ : label -> type1
        op type2?_ : label -> type2
        op type3?_ : label -> type3
        var t1 : type1
        var t2 : type2
        var t3 : type3
        eq type1?(t1 &  t2 & t3) = t1 .
        eq type2?(t1 & t2 &  t3) = t2 .
        eq type3?(t1 & t2 &  t3) = t3 .
        eq (t1 = t1 ) = true .
        eq (t2 = t2 ) = true .
        eq (t3 = t3 ) = true . }
```

Figure 3. The module defining a Label in CafeOBJ notation

Content and specifies what actions are allowed on contents as defined by the OMA specification (play, display, print, export, execute). These are specified as constants of the action sort. In addition it defines the concept of a permission as a binary operator on the sorts of action and cont.

3. *Request*, defines the visible sorts of req and reqErr that specify a user request and an invalid user request respectively. It imports the modules Permission and Content. A request is specified as a pair of content and an action.
4. *Colors*, defines the visible sort color and the constants black and white of that sort.
5. *Label*, defines the visible sorts of type1, type2, type3 and label. As type1 the constants simple, complex are defined. Next as type2 the constants count, datetime, true are defined. Finally as type3 we define the constants once, many. A label is defined by the operator _&_&_ that takes as input an element of type1×type2×type3. Also three operators that take as input a label and return the corresponding type are defined; type1?, type2?, type3? . The CafeOBJ code for this module is given in figure3.
6. *SET*, is a parameterized module that defines the visible sorts of Elt and Set. These correspond to an abstract arbitrary data type that will act as elements for a set and a set of such elements respectively. The module is parameterized meaning that it can be instantiated with another module, which will instantiate the elements, so as to specify the set of a specific type of elements.
7. *Constraint, ConstraintSet* are two modules specifying the constraints defined by OMA and a set of such constraints respectfully.
8. *ConstraintPermission, SetofCP*, are two modules specifying a

| Observers Signature | Description |
|---|---|
| `validsublic : Lsys subLic -> Bool` | Returns true if the constraints of the sublicense are met at the given state |
| `bop validCP : Lsys cPerm -> Bool` | Returns true if the constraints of the constraint permission set are met at the given state |
| `bop installed : Lsys -> licSet` | Returns the set of installed licenses |

Table 3. Observers of the component OTS

constraint permission set as a set of constraints when met allows a set of permissions and a set of such elements respectfully.

*Sublicense*, defines the visible sort `subLic` that specifies the concept of a sublicense. A set of constraints that when it is met allows a set of constraint permission sets.

9. *License*, defines the visible sort `lic` that instantiates the module `SET` by defining a set with sub-licenses as elements. This corresponds to the definition of a license in OMA REL; a set of sub-licenses.
10. *LicSet*, defines the visible sort `licSet`, that by instantiating the module `SET` again defines a set of licenses.

### 4.4.2 OTS Model and its Specification in CafeOBJ

Next we define the hidden sorts for our OTS. Each such sort defines the state space of an abstract machine. In our specification we require two such sorts. The first sort, `Lsys`, specifies a system that consists of a set of licenses that can deplete a whole sublicense or a constraint permission set after a relative request. This sort is defined in module LOTS that imports the module LicSet. The observation operators it contains are presented in table 3.

In this OTS all the installed licenses have their constraints met at the initial state of the system. This is defined by the equation: `ceq validsublic(initl, SL1) = true if (SL1 /inL installed(initl))`. Where `SL1` is a CafeOBJ variable denoting an arbitrary sub-license, `/inL` is an operator that checks if a sub-license is contained in a set of licenses and `initl` is a hidden sort constant denoting an arbitrary initial state of the OTS.

The transitions of the OTS can be seen in table 4. Assuming that S, CP1 are variables defining an arbitrary state and an arbitrary constraint permission respectively while SL1, SL2 are variables defining arbitrary sub-licenses, the equations defining the first transition can be seen in table 5.

| Transition Signatures | Description |
|---|---|
| `bop depleteSL : Lsys subLic -> Lsys` | Models the transition that occurs when the whole of the given license becomes depleted |
| `bop depleteCP : Lsys cPerm -> Lsys` | Models the transition that occurs when a constraint permission set of a sublicense becomes depleted. |

Table 4. Transitions of the first OTS

```
eq installed(depleteSL(S , SL1)) = installed(S) .
ceq validsublic(depleteSL(S , SL1) , SL2) = validsublic(S , SL2)
if not (SL1 = SL2) .
ceq validsublic(depleteSL(S , SL1) , SL2) = false  if  (SL1 =
SL2) .
ceq validCP(depleteSL(S , SL1) , CP1) = validCP(S , CP1) if not
( CP1 /inCP3 SL1) .
ceq validCP(depleteSL(S , SL1) , CP1) = false if  ( CP1 /inCP3
SL1) .
```
Table 5. Equations defining the depleteSL transition.

Our goal is to specify a DRM agent that uses the new algorithm. So this OTS does not suffice. We will use this OTS as a component object of our new OTS that specifies our system. This OTS defines a new state space denoted by the hidden sort `sys`. In order to derive the state of the component object from the composite object we use *projection operators*. These are special kind of operators that given a state of the composite system return a state of one of the components. Since in this specification we only have one component we only define one projection: `bop license_ : sys -> Lsys`. For example assuming that init is a CafeOBJ constant that denotes the initial sate of the composite OTS, we derive the state of the component system with the following equation, stating that the component object will be at its initial state as well: `eq license(init) = initl` .

In order to specify the behavior of a system that implements the proposed algorithm for an arbitrary set of installed licenses the observers of table 6 where required. To conclude the definition of our OTS we must define the transitions that will take place, the conditions under which these rules change the state of the system successfully, and finally what each observer observes in a state that is derived from an arbitrary state after applying these rules. The transitions used are shown in table 7. Where in the description with subl we denote a sublicense such that for user request r $sat(subl, r)$ holds and with cp we denote a constraint permission such that $cp \in subl$ and $sat(cp, r)$. Also in table 7 we refer to the following property as Prop5: cp is the only constraint permission such that $sat(cp, r)$ or for all other $cp' \in subl'$ such $sat(cp', r)$ holds then $(times(subl') = once \land comp(subl') = Complex) \lor (times(cp') = once \land comp(cp') = Complex)$.

Table 6. Observers of the composite OTS

| Observer Signature | Description |
| --- | --- |
| `bop licIns : sys -> licSet` | Returns the set of installed licenses. |
| `bop useReq : sys -> reqErr` | Returns the request of the user at the given state, if it exists. Else returns an error constant |
| `bop best : sys  -> lic` | Returns the license the algorithm selects for the user request |
| `bop color : sys perm ->  color` | Returns the color of the given permission |
| `bop possLic : sys -> licSet` | Returns a set of licenses that contain the user request in a constraint permission set that is |
| | not labeled as Once or is labeled as Simple. |
| `bop finalLic : sys -> licSet` | Returns a set of licenses that belong to possLic and the labels of the sublicenses where the request belongs are not labeled as Once or are labeled as Simple. |
| `bop allowed : sys -> permSet` | Returns the set of permissions allowed by the installed licenses initially. |

As an example of the specification process with the OTS/CafeOBJ system we will analyze the code specifying the `use3` transition in table 8. The numbers in the beginning of each line are not part of the code and are used as reference. In equations 1 to 2 we define the effective condition for the transition rule. `Find4` is an operator that takes as input a request and a license and returns the sub-license the request belongs to. `label?` is another operator that takes as input a sub-license and returns its label. The next operator `#_` takes as input a set and returns the number of elements in it. Equation 3 defines a mapping from the state of the system to the state of the component OTS through the projection operator. This equation states that new state of the component OTS will be attained from the old by depleting the sublicense containing to the user request. This is achieved by reference to the transition of the component object `depleteSL`. Lines 4 to 5 state, that the set of installed licenses remains unchanged by this transition, and that we no longer have a pending user request.

| Transition Signatures | Description |
|---|---|
| `bop request : reqErr sys -> sys` | Denotes that a user makes a request |
| `bop choose  : sys -> sys` | Models the transition where the DRM agent selects a license for the user request |
| `bop use1    : sys -> sys` | Models the transition where the algorithm selects and uses a license in which $times(cp) = Once, times(subl) = Many$ and prop5 holds. |
| `bop use3    : sys -> sys` | Models the transition where the algorithm selects and uses a license in which $times(subl) = Once$ and prop5 holds. |
| `bop use4    : sys -> sys` | Models the transition where the algorithm selects and uses a license for which $times(cp) = Once, times(subl) = Many$ and prop5 does not hold. |
| `bop use5    : sys -> sys` | Models the transition where the algorithm selects and uses a license in which $times(subl) = Once$ and prop5 does not hold. |

Table 7. Transitions of the composite OTS

```
1.  op c-use3 : sys -> Bool
2.  eq c-use3(S) =(not(useReq(S) = null) and (not (best(S) =
    emptyLic))) and (type3?(label? (find4(useReq(S), best(S)))) =
    once) and ((# build2(useReq(S) , licIns(S) ,license(S))== 1)
    or ( (possLic(S) = emptyLic) or  (finalLic(S)= emptyLic)))  .
3.  ceq license(use3(S))=depleteSL(license(S),find4 (useReq (S),
    best(S))) if c-use3(S) .
4.  ceq   licIns(use3(S)) = licIns(S) if c-use3(S) .
5.  ceq   useReq(use3(S)) = null if  c-use3(S) .
6.  ceq    color(use3(S) , P) = black if (P = perm3?(useReq(S)
    ,find3(useReq(S),best(S)))) and (P /in allowed(use3(S)))  and
    c-use3(S) .
7.  ceq   color(use3(S),P)=  black  if  (belong3?(makeReq(P)   ,
    skolem(P))  and  (skolem(P)  /inCP2 best(S))  and  (P /in
    allowed(use3(S))) and c-use3(S)) .
8.  ceq  color(use3(S),P)=color(S,P)  if  belong3?( makeReq(P),
    skolem(P))and not  (skolem(P)/inCP2 best(S))and(P/in  allowed
    (use3(S))) and c-use3(S)   .
9.  ceq color(use3(S),P)= color(S,P) if not(P = perm3?(useReq(S)
    , find3(useReq(S),best(S)))) and ( not(belong3?(makeReq(P) ,
    skolem(P))  and  (skolem(P)  /inCP2  best(S))))  and  (P  /in
    allowed(use3(S))) and c-use3(S)   .
10. ceq finalLic(use3(S))= finalLic(S) if c-choose(S) .
11. ceq best(use3(S)) =  emptyLic if c-use3(S) .
12. ceq allowed(use3(S)) = allowed(S) if c-use3(S) .
13. ceq use3(S) = S if not c-use3(S)  .
```
Table 8. Definition of the use3 transition in CafeOBJ notation

Equations 6 to 9 define the coloring of arbitrary permission elements. The first equation states that a permission element is colored black if it equals the user request and belongs to the license the algorithm selected. The operator `perm3?` takes as input a user request and a constraint permission and returns the permission that matches the request. Also `find3` is an operator similar to `find4` only that it returns the constraint permission set containing the request. Equation 7 uses Skolemization (via the operator skolem_) to denote that a permission p gets colored black if there exist a constraint permission set cp, in the sub-license of the selected license that contains the user request that becomes depleted such that $p \in cp$. Skolemization is required because there is no other way to denote existential operators in CafeOBJ. The operator `belong3?`, returns true if a permission belongs to a constraint permission set.

| Liveness Property in CafeOBJ Notation | Liveness Property in Natural Language |
| --- | --- |
| `op lto : sys perm  -> Bool`<br><br>`eq lto(S, P) = ((color(S , P) = white) /\ (P /in allowed(S)))  \|--> (color(S, P) = black ) .` | *if a permission right belongs to the installed licenses of the DRM system and is Colored White that leads to it being Colored Black* |

Table 9. Liveness Property to Prove

The operator `_/inCP2_` checks if a constraint permission set belongs to a license. Equation 8 states that the color of a permission should remain unchanged, if it belongs to a constraint permission set that does not belong to the sublicense of the selected license that contains the request. Finally equation 9 states that the color should again remain unchanged if a permission is does not match the request and does not belong to a constraint permission set in the license that the algorithm has chosen. Equations 10 to 13 define that the elements of the finalLic array should remain unchanged, that the request is fulfilled so there is no longer a license selected, that the permissions originally allowed by the licenses remain unchanged and that finally the state of the OTS does not change if effective condition is not met.

### 4.5 Verification of the algorithm

The first step of the verification is to define the property we wish to prove in a module that imports OTSLOGIC, usually called LTO. The desired property can be seen in table 9, where S is a CafeOBJ variable denoting a state of the system and P variable denoting an arbitrary permission.

The next step is to ask CafeOBJ to reduce if the property holds in a script (called proof score). If CafeOBJ returns true, means that the property is proved. This is achieved using the `red` command that reduces a term based on the specification and by replacing the variables with arbitrary constants:

```
open LTO
red  lto(s,p) .
close .
```

In most cases, CafeOBJ will return an expression different than true or false. This means that it cannot continue with the rewriting and we must feed it with additional equations or lemmas to help it.

In the case of verifying liveness properties it is helpful to look at the deduction rules defined in OTSLOGIC and see if it is possible to break down the property in to *leads-to, ensures* and *invariant* properties. The desired property here was proved using two ensure properties ens1, ens2 and CafeOBJ returned true for the following proof passage:

```
open LTO
red (ens1(s , p ) /\ ens2(s , p )) =>  lto(s , p) .
close .
```

The two ensure properties can be seen in table 10. So now in order to complete our verification we must prove these two properties.

| The first *ensures* property |
|---|
| ```
eq ens1(S , P) =  (((makeReq(P) = useReq(S)) \/
(belong3?(makeReq(P) , find3(useReq(S) , best(S))) /\
(type3?(labelCP?(find3(useReq(S), best(S) ))) = once)
/\(not(type3?(label?(find4(useReq(S),best(S)))) = once))   /\  (
(# build2(useReq(S) , licIns(S),license(S)) == 1)
\/(possLic(S)= emptyLic)\/ (finalLic(S)= emptyLic)))
\/(belong3?(makeReq(P),skolem(P)) /\ (skolem(P) /inCP2 best(S)
/\ ~(best(S) = emptyLic))/\(type3?(label?(find4(useReq(S) ,
best(S)))) = once) /\ ((# build2(useReq(S) , licIns(S),
license(S))== 1)  \/ ( possLic(S) = emptyLic) \/ (finalLic(S) =
emptyLic))))/\(P /in allowed(S))) ensures (color(S , P ) = black
).
``` |
| The second *ensures* property |
| ```
eq ens2(S ,P ) = (((color(S , P) = white) /\ (P /in allowed(S)))
)   ensures  (((makeReq(P) = useReq(S)) \/  (
belong3?(makeReq(P) , find3(useReq(S) , best(S)))  /\
(type3?(labelCP?(find3(useReq(S), best(S) ))) = once)  /\
(not(type3?(label?(find4(useReq(S) , best(S)))) = once))  /\  (
(# build2(useReq(S) , licIns(S),license(S)) == 1)   \/
(possLic(S) = emptyLic)   \/   (finalLic(S)= emptyLic)
) ) \/  (    belong3?(makeReq(P) , skolem(P)) /\ (skolem(P)
/inCP2 best(S)   /\ ~(best(S) = emptyLic) )   /\
(type3?(label?(find4(useReq(S) , best(S)))) = once)   /\
(# build2(useReq(S) , licIns(S), license(S))== 1)   \/(
possLic(S) = emptyLic)\/  (finalLic(S) = emptyLic)  ))) /\ (P
/in allowed(S)) )    .
``` |

Table 10. The first two lemmas required for the verification

For each ensures property we first must prove the *unless case* and the *eventually case*. Here we will only present the verification of the *unless case* of the first ensures property for one transition rule. In total 7 lemmas where required to complete the verification of these cases and they are provided in table 11. Three of them are lemmas on the states of the OTS while the rest are lemmas on the data-types.

We will present here the subcase for the verification of the `use4` transition. We ask CafeOBJ to check this using the following proof score:

| CafeOBJ notation | Natural Language Notation |
|---|---|
| ```
eq inv2(S,P)=
((not(P=perm3?(useReq(S),
find3(useReq(S),best(S))))) and
not(useReq(S)= null) and not(P =
err) )                implies not
( makeReq(P) = useReq(S)) .
``` | If a permission element is not the equal to the outcome of finding the permission of license chosen by the algorithm as the best to use and it is not an error (null) and there is a valid user request pending then that permission does not equal the user request. |

| | |
|---|---|
| `eq inv3(S , P) =  (P /in allowed(init)) implies (P /in allowed(S)) .` | If a permission element belongs to the set of initially allowed permission by the installed licenses then it keeps belonging to that set in all stages. |
| `eq inv4(S , P) =  ((P = perm3?(useReq(S),find3(useReq(S) , best(S)))) and not(useReq(S)= null) and not(P = err) ) implies  (makeReq(P)= userReq(S)).` | If a permission element is equal to the outcome of finding the permission of license chosen by the algorithm as the best to use and it is not an error (null) and there is a valid user request pending then that permission is equal the user request. |
| `eq inv5(R ,CP , P)  = ((P = perm3?(R ,CP) ) and not(R= null )) implies ( R = makeReq(P)) .` | If a permission element is equal to the out-come of matching a request in a constraint permission and it is not null then the permission element equals the request. |
| `eq inv6(R,PS,P)   = ((P = perm2?(R ,PS) ) and not(R = null ) and not (perm2?(R ,PS) = err)  )   implies ( R = makeReq(P) ) .` | If a permission element is the match for a request in a permission set and it is valid and the request is valid that implies that the request matches the permission element |
| `eq inv7(R , P) = not(belong1?(R ,P) and not(makeReq(P) = R )).` | if a request does not equal to a permission element then transforming that permission into a request does not match it. |

Figure 7. Lemmas required for the verification

```
open USTEP
op p : -> perm   .
op cp : -> cPerm   .
op r : -> req .
eq s' = use4(s).
red ustep1(s , p ) .
close
```

CafeOBJ returns neither true nor false here. Instead it returns an expression from which we can conclude that the machine cannot deduce whether the effective condition for this rule holds. So we result to case-splitting to assist the machine with the rewriting, the first case that denotes that the effective condition does not hold is defined as:

```
open USTEP
op p : -> perm   .
op cp : -> cPerm   .
op r : -> req .
eq s' = use4(s).
```

```
-- c-use4(s) = false .
eq useReq(s) = null .
eq best(s) = emptyLic .
eq (type3?(labelCP?(find3(useReq(s), best(s) ))) = once )
= false .
eq  type3?(label?(find4(useReq(s) , best(s)))) = once .
eq (#build2(useReq(s),licIns(s),license(s))== 1) = true .
eq possLic(s) = emptyLic .
eq finalLic(s) = emptyLic .
red  ustep1(s,p) .
close
```

CafeOBJ returned true here so this sub-case is proved. What is left to check is its symmetrical case, i.e. when the effective condition holds for the transition rule:

```
open USTEP
op p : -> perm   .
op cp : -> cPerm   .
op r : -> req .
eq s' = use4(s).
-- c-use4(s) = true .
eq (useReq(s) = null) = false .
eq (best(s) = emptyLic) = false  .
eq type3?(labelCP?(find3(useReq(s), best(s) ))) = once  .
eq (type3?(label?(find4(useReq(s) , best(s)))) = once) =
false  .
eq  (# build2(useReq(s) ,  licIns(s),license(s))==  1)  =
false .
eq (possLic(s) = emptyLic) = false .
eq (finalLic(s) = emptyLic) = false  .
red  ustep1(s,p) .
close
```

CafeOBJ returns again neither true nor false. Instead it returns an expression from which we can conclude that it cannot rewrite the predicate color(s,p) = black to true or false. So we once must help with the rewriting by splitting this case as:
```
open USTEP
op p : -> perm   .
op cp : -> cPerm   .
op r : -> req .
eq s' = use4(s).
-- c-use4(s) = false .
eq (useReq(s) = null) = false .
eq (best(s) = emptyLic) = false   .
eq type3?(labelCP?(find3(useReq(s), best(s) ))) = once  .
eq (type3?(label?(find4(useReq(s) , best(s)))) = once) =
false   .
```

```
eq (# build2(useReq(s) , licIns(s),license(s))== 1) =
false .
eq (possLic(s) = emptyLic) = false .
eq (finalLic(s) = emptyLic) = false  .
-- CASE SPLITING
eq color(s ,p) = black .
red  ustep1(s,p) .
close
```

CafeOBJ returns true for this case but again neither true neither false for the symmetrical case. So again we conclude what predicate cannot be further rewritten using the given equations. Continuing in the same approach we reach the case denoted by the predicates: `c-use4(S)/\~color(s,p) = black/\~(p /in allowed(s)/\perm3?(useReq(s),find3(useReq(s),best(s)))=p`. where CafeOBJ has returned true for all the symmetrical sub-cases. Now we could continue and split the case more and see what result we would get, but since this is a computer human interactive procedure we conclude to the result that not all of these equations may hold simultaneously in our specification. So we produced here invariant2 and used it as a lemma to discard this case as:

```
open USTEP
op p : -> perm   .
op cp : -> cPerm  .
op r : -> req .
eq s' = use4(s).
-- c-use4(s) = false .
eq (useReq(s) = null) = false .
eq (best(s) = emptyLic) = false   .
eq type3?(labelCP?(find3(useReq(s), best(s) ))) = once  .
eq (type3?(label?(find4(useReq(s) , best(s)))) = once) =
false   .
eq (# build2(useReq(s) , licIns(s),license(s))== 1) =
false .
eq (possLic(s) = emptyLic) = false .
eq (finalLic(s) = emptyLic) = false   .
-- CASE SPLITING
eq (color(s ,p) = black) = false   .
eq (p /in allowed(s) ) = true .
eq (perm3?(useReq(s) , find3(useReq(s) , best(s))) = p) =
false .

red inv2(s ,p) implies  ustep1(s,p) .
close
```

The above proof score returned true as well so this concludes the proof for this transition rule. Of course this procedure continued until all the transitions where proven to preserve the unless inductive step, and all the lemmas used where in turn proved to hold for every reachable state of our OTS (since they are all invariant lemmas). Finally eventually step a witness transition was found, use1 in this case.

This concludes the proof of the first ensure property. The same procedure was followed successfully for the second eventually property as well and this, together with the proofs of the lemmas, concludes the proof of the leads to property.

In this section we have provided a formal proof that our algorithm does not case this loss of rights, by providing a proof as a liveness property in the OTS/CafeOBJ framework.

## 5. Conclusions and Work in Progress

We have redesigned the OMA License Choice Algorithm by simply introducing labels on some parts of the licenses. This can be easily done by tweaking the way licenses are created since these are written in XML. The extra computational cost has been shifted to the creation of licenses. We do not believe that this will cause any problems since these licenses are written by computers with much higher computational power. We also believe that it will be easier to create automated tools that take as input a license written in OMA REL and give as output licenses containing these labels. In addition, we have produced some case studies through an implementation of the algorithm in Java that we believe demonstrates the benefits of this algorithm over the existing one, as well as provide hints that it behaves in a correct manner. Of course no matter how many case studies one does it does not provide a formal proof of the correctness of the algorithm being tested. To this end we have created a specification of this new algorithm as an Observation Transition System (OTS) written in CafeOBJ terms. This allowed us to formally prove that the algorithm we provide here does not cause the loss of permissions rights discussed. The property we verified was a liveness property and to our knowledge this is one of the first applications of this technique in a real life problem. We must point out that although the algorithm studied and redesigned is originally designed to address the decision problem, of what licenses to use when multiple licenses refer to the same content, for the mobile environment it could be tweaked to address the same problem in different environments since all of these systems use licenses written in RELs that have similar constraints and an ordering on these constraints is the bases of these kind of algorithms. This work is in the spirit of our previous efforts in the formal verification of mobile systems and algorithms [8, 9 and 10].